\documentclass[aps,physrev,preprint,superscriptaddress]{revtex4-2} %

\usepackage[french, english]{babel}
\usepackage{amsmath}
\usepackage{graphicx}
\usepackage[colorlinks=true, allcolors=blue]{hyperref}
\usepackage[utf8]{inputenc}
\usepackage{amsfonts}
\usepackage{array}
\usepackage{gensymb}
\usepackage{float}
\usepackage[version=4]{mhchem}
\usepackage{soul}

\begin{document}


\title{Fracture initiation in silicate glasses via a universal shear localization mechanism} %



\author{Matthieu Bourguignon}
\affiliation{Soft Matter Sciences and Engineering, ESPCI Paris, PSL University, CNRS, Sorbonne University, 75005 Paris, France.}

\author{Gustavo Alberto Rosales-Sosa}
\author{Yoshinari Kato}
\affiliation{Nippon Electric Glass, 7-1, Seiran 2-Chome, Otsu, 520-8639, Shiga, Japan.}

\author{Bruno Bresson}
\affiliation{Soft Matter Sciences and Engineering, ESPCI Paris, PSL University, CNRS, Sorbonne University, 75005 Paris, France.}

\author{Hikaru Ikeda}
\author{Shingo Nakane}
\affiliation{Nippon Electric Glass, 7-1, Seiran 2-Chome, Otsu, 520-8639, Shiga, Japan.}

\author{Gergely Molnár}
\affiliation{CNRS, INSA Lyon, LaMCoS, UMR5259, 69621 Villeurbanne, France}

\author{Hiroki Yamazaki}
\affiliation{Nippon Electric Glass, 7-1, Seiran 2-Chome, Otsu, 520-8639, Shiga, Japan.}

\author{Etienne Barthel}
\affiliation{Soft Matter Sciences and Engineering, ESPCI Paris, PSL University, CNRS, Sorbonne University, 75005 Paris, France.}

\date{\today}

\begin{abstract}
Shear bands lie at the root of fracture initiation in bulk metallic glasses and amorphous polymers. For silicate glasses, in contrast, studies have largely emphasized permanent volumetric strain, commonly referred to as densification. Here we systematically investigate indentation-induced fracture in two distinct families of aluminoborosilicate glasses. The results demonstrate that plastic shear flow plays a decisive role in governing fracture initiation. In addition, molecular dynamics simulations reveal a pronounced composition dependence of softening associated with plastic shear flow, closely mirroring the experimentally observed propensity for strain localization. We conclude that silicate glasses conform to a universal  pattern of rupture initiation governed by localization of shear-deformation, aligning with a broad range of amorphous materials, including bulk metallic glasses and glassy polymers.
\end{abstract}


\maketitle


\section{Introduction} %

Despite decades of intensive research, a comprehensive understanding of the rupture properties of silicate glasses remains a significant challenge~\cite{Wondraczek22}, constraining mitigation strategies. A range of conceptual frameworks has been advanced, including the tailoring of network connectivity~\cite{Januchta19}, the notion of optimal topological rigidity~\cite{Wondraczek11} (often illustrated by highly connected networks such as silicon oxynitrides or carbides~\cite{Rouxel_2021}) or, from a quite different perspective, the ideas of nanoscale ductility~\cite{Wondraczek11} or adaptive network structures~\cite{Januchta17a}. These approaches however, offer limited predictive capability and have yet to establish clear, quantitative relationships between glass composition, underlying deformation mechanisms, and macroscopic mechanical performance.

One of the difficulties is that, in very practical terms, the fracture sensitivity of silicate glasses is usually probed by indention, indeed a technique of choice for characterizing the non-linear deformation, damage and fracture processes in brittle materials~\cite{Marshall2015}. In particular, indentation cracking provides a convenient and quantitative measure of crack resistance (CR), the resistance to indentation-induced damage~\cite{Wada74, Lawn80}. In fact, although CR, defined as the threshold load required to initiate median or radial cracks beneath a Vickers indenter tip, may seem somewhat primitive at first glance, it has proven to be highly discriminating for silicate glasses~\cite{Kato10}.

The strain field under the tip is complex, but extensive studies over more than half a century~\cite{Lawn75,Chiang1982,Davis20,RosalesSosa25} have provided a robust foundation for a general understanding of the ensuing damage and fracture processes. Silicate glasses, however, undergo irreversible volumetric deformation, or densification~\cite{Ernsberger68} which reduces indentation-induced stresses and can therefore partially suppress crack initiation and propagation~\cite{Yoffe1982,Cook90}. As a result, over the past decades, considerable effort has been devoted to quantifying the role of densification in decreasing the cracking sensitivity of silicate glasses~\cite{Rouxel15,Barlet15}. Notably, low atomic packing density and a low Poisson’s ratio, both linked to enhanced densification, have been reported to correlate with improved crack resistance~\cite{Yoshida20, Rouxel_2021}. 

In contrast to this densification-oriented perspective, an alternative framework was developed in the 1980s, highlighting localized shear deformation as the primary mechanism governing fracture initiation. Experimental studies showed that cracks often originate at discrete shear events, referred to as shear faults or slip events, which appear during loading and serve as deformation-induced crack nuclei~\cite{Hagan79,Lawn83,Lathabai91}. This line of investigation, however, has received comparatively little attention in recent years, with only one notable exception~\cite{Gross18}.

To clarify the relative contributions of the different mechanisms proposed for fracture initiation in silicate glasses in relation to their composition and structure, we conducted a systematic investigation of indentation cracking across two families of glass compositions spanning a broad range of crack resistances. One of these families comprises Ca alumino-boro-silicate glasses: the influence of silicon substitution by boron on crack resistance is well established and has been extensively studied~\cite{Kato10a}, although a complete understanding of the mechanisms underlying boron-induced strengthening remains elusive. To broaden the scope of our conclusions, we also examined the effect of network modifier field strength, substituting Mg for Ca, another well-known strategy for drastically improving crack resistance~\cite{Osada20, Fry23}.

Within these compositional ranges, we find compelling evidence that crack resistance correlates strongly with the morphology of shear bands, while exhibiting no significant dependence on Poisson’s ratio or on the densification behavior. This finding demonstrates that, in silicate glasses, the intrinsic resistance to localized shear faulting controls the macroscopic cracking response under sharp-contact loading. Shear flow and the associated material instabilities thus emerge as a unifying paradigm of fracture behavior in amorphous materials, one that silicate glasses, despite their capacity for densification, share with  \textit{e.g.} metallic glasses and glassy polymers.

\section{Results}
Eighteen glass samples were melted (see Supplemental Material, section S~I for more detail on the experimental procedures) with the nominal composition \((15-y)\text{CaO} \cdot y\text{MgO} \cdot 15\text{Al}_2\text{O}_3 \cdot x\text{B}_2\text{O}_3 \cdot (70-x)\text{SiO}_2\) (CMABS glasses). The boron content (x) was varied as 0, 5, 15, and 25~mol\% (Table~S1). For compositions with $x$ = 5 and 15~mol\%, CaO was progressively substituted by MgO, with $y = 0$ (CABS), 2.5, 5, 7.5, 10, 12.5 and 15 (MABS)~mol\% (Table~S2). In addition, reference glasses previously reported by Kato et al.~\cite{Kato10a} were also prepared for comparison.

\subsection{General mechanical properties}
Key mechanical properties were systematically characterized and are presented in Fig.~\ref{fig:MechanicalProperties}, with numerical values provided in the Supplemental Material, section S~I. Young’s modulus $E$ and shear modulus $G$ were measured with the resonance method, hardness $H$ with microindentation under a load of 100~gf. 

\begin{figure}[] %
\centering
a)\includegraphics[width=0.32\textwidth]{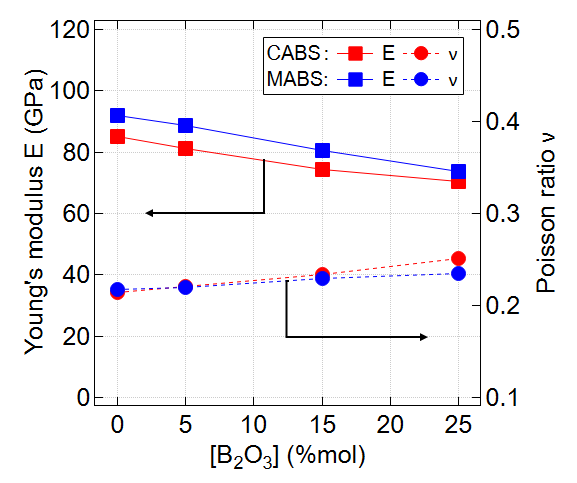}
b)\includegraphics[width=0.3\textwidth]{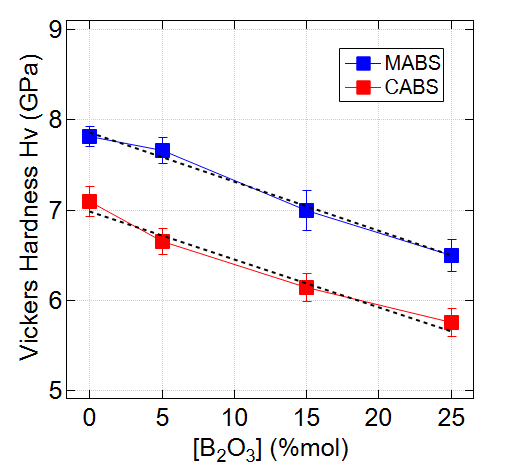}
c)\includegraphics[width=0.3\textwidth]{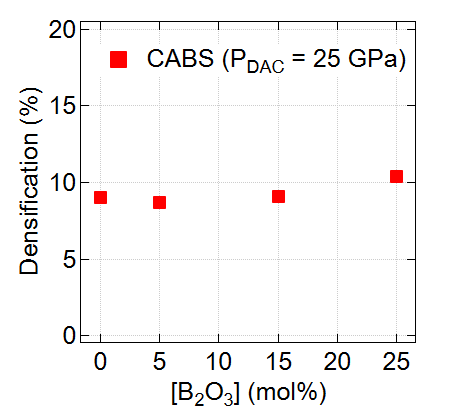}
d)\includegraphics[width=0.32\textwidth]{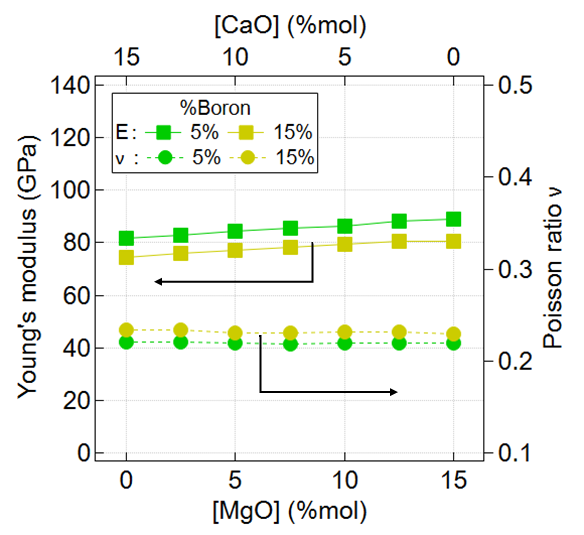}
e)\includegraphics[width=0.3\textwidth]{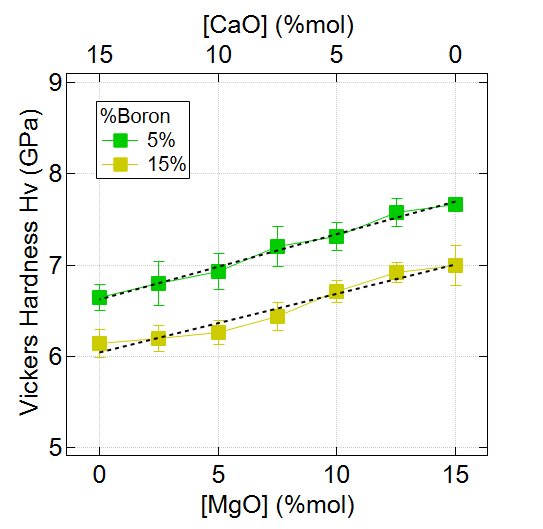}
f)\includegraphics[width=0.3\textwidth]{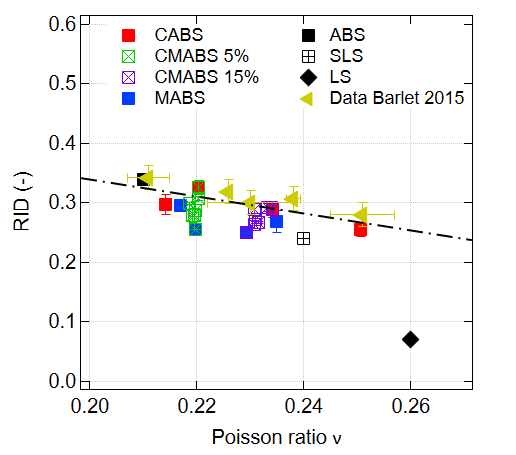}
\caption{(a-c) single alkaline earth (Ca or Mg) ABS glasses with varying B$_2$O$_3$ content: a) Young's Modulus $E$ and Poisson's ratio $\nu$; b) micro-indentation Vickers hardness H$_V$; c) densification (CABS only) at 25 GPa (Multi Anvil Cell)~\cite{Gomes25} - (d-e) mixed alkaline earth ABS glasses (5\% or 15\% B$_2$O$_3$) as a function of Mg vs Ca content: d) Young's Modulus $E$ and Poisson ratio $\nu$; e) micro-indentation Vickers hardness H$_V$ - (f) all glass compositions: RID as a function of Poisson ratio including some (LS, SLS, ABS) glasses from Kato \textit{et al} \cite{Kato11} and from Barlet \textit{et al.} \cite{Barlet15}. Error bars are not shown if smaller than the marker size.}
\label{fig:MechanicalProperties}
\end{figure}

Overall, the property evolutions were moderate across the present compositional changes. Incorporating 25 mol\% $\text{B}_2\text{O}_3$ led to a 20\% reduction in both Young's modulus and hardness. Conversely, substituting $\text{Mg}^{2+}$ for $\text{Ca}^{2+}$ resulted in a marginal 5\% increase in $E$ and a 10\% increase in $H$. Regarding Poisson's ratio, the 25 mol\% $\text{B}_2\text{O}_3$ addition caused a 20\% increase in the CABS system and a 10\% increase for the MABS system, while the $\text{Mg}^{2+}$ for $\text{Ca}^{2+}$ substitution had negligible effect on $\nu$.

The densification behavior of the glasses was also investigated. For the CABS series, densification under hydrostatic pressure was measured up to 25~GPa using a diamond anvil cell (DAC). A densification of approximately 10 \% was observed, with no measurable dependence on boron content (Fig.~\ref{fig:MechanicalProperties}~c). For all compositions, indentation-induced densification was evaluated via the recovery of indentation depth (RID) method \cite{Yoshida05}, following established procedures \cite{Yoshida05,Kato10,Fry23}. Vickers indentation depths, $d_{\text{before}}$ and $d_{\text{after}}$, were measured before and after annealing at 0.9 $T_g$ for 2 h, and the RID was calculated as

\begin{equation}
RID = \frac{d_{\text{before}} - d_{\text{after}}}{d_{\text{before}}}.
\label{eqn:RID}
\end{equation}

The RID values are approximately 0.3 and exhibit minimal variation across compositions (Fig.~\ref{fig:MechanicalProperties}f). A correlation with Poisson’s ratio is observed, characterized by a slight decrease in RID with increasing Poisson’s ratio, consistent with previous observations \cite{Rouxel08}. Overall, these results agree well with those reported by Barlet et al. \cite{Barlet15} for glasses of similar compositions.

\subsection{Crack resistance} %
To determine the crack resistance (CR), glass samples were indented using a Vickers indenter under controlled environmental conditions (25 °C and 30 \% relative humidity). Radial crack formation at the indentation corners was systematically recorded. The applied load was incrementally increased from 100 gf to 6 kgf, with 40 indentations performed at each load. The probability of crack initiation $P_{\mathrm{CI}}$ was calculated as the total number of observed radial cracks $n_{\mathrm{RC}}$ over 4, the total number of indentation corners. The crack resistance, CR, was defined as the applied load corresponding to $P_{\mathrm{CI}} = 50\%$.

\begin{figure}[] %
\centering
a) \includegraphics[width=0.4\textwidth]{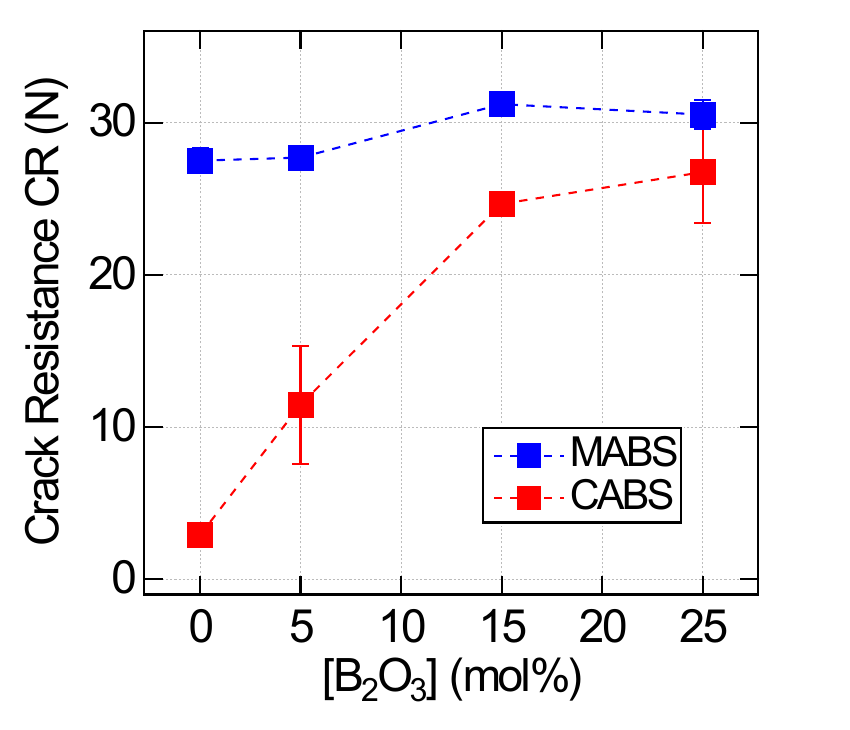}
b) \includegraphics[width=0.4\textwidth]{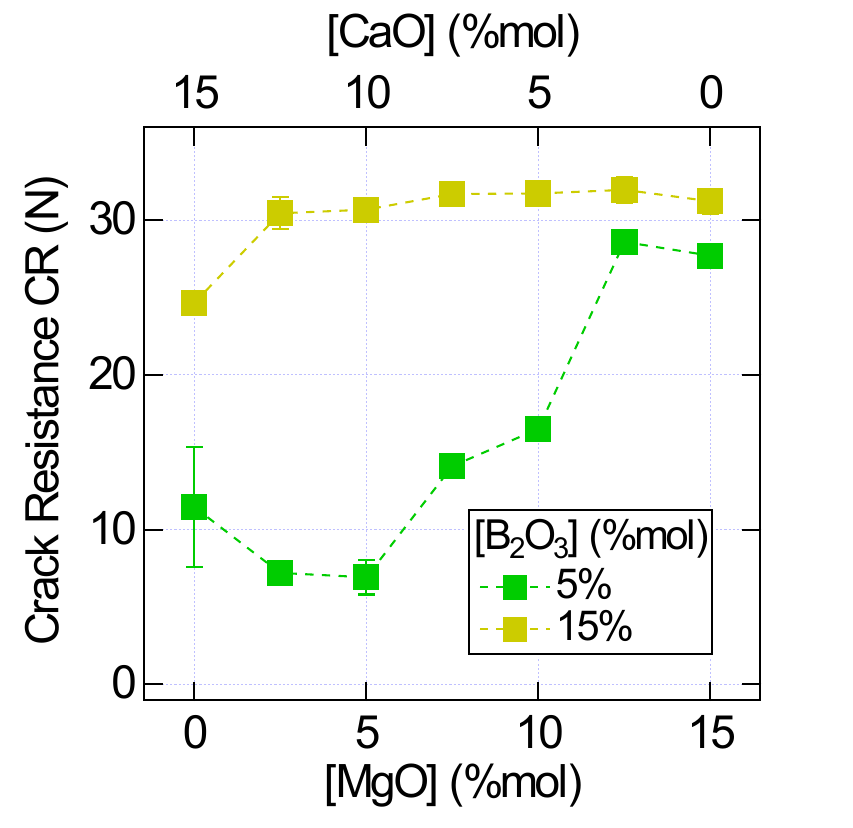}
c)\includegraphics[width=0.4\textwidth]{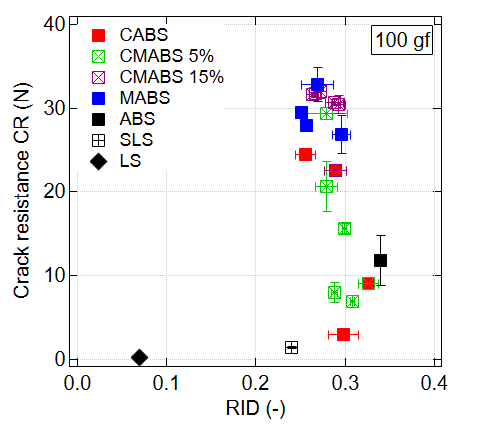}
d)\includegraphics[width=0.4\textwidth]{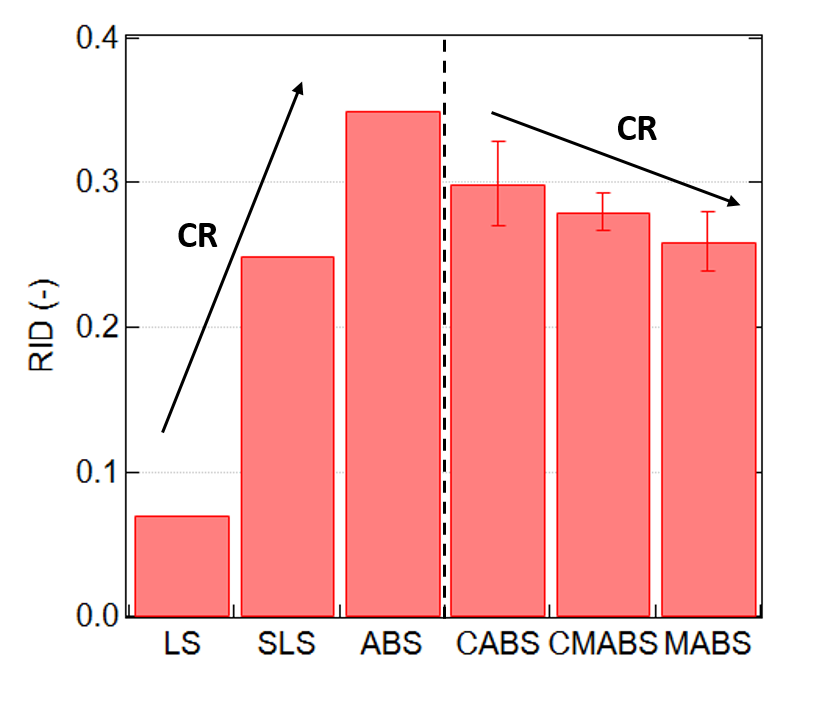}
\caption{(a-b) crack resistance $CR$ of (a) single alkaline earth network modifier ABS glasses and (b) mixed alkaline earth network modifiers ABS glasses. If no error bar is shown, it is less than the markers size - (c) crack resistance ($CR$) as a function of recovered indentation depth ($RID$) - (d) recovered indentation depth for different silicate glasses compositions including some (LS, SLS, ABS) from Kato et al \cite{kato2011load}. Arrows show how crack resistance increases with compositions. }
\label{fig:CR}
\end{figure}

The measured CR values for all compositions are plotted in Fig.~\ref{fig:CR}. In contrast to the other mechanical properties, CR exhibited a strong compositional dependence. For the CABS series (Fig.~\ref{fig:CR} a), increasing the $\text{B}_2\text{O}_3$ content from 0 to 25 mol\% resulted in a significant ten-fold increase in CR. Similarly, when the $\text{B}_2\text{O}_3$ content was fixed at 5 mol\% (Fig.~\ref{fig:CR} b), substituting $\text{MgO}$ for $\text{CaO}$ initially caused a slight decrease in CR between 0 and 5 mol\% $\text{MgO}$, followed by a substantial three-fold increase between 5 mol\% and 15 mol\% $\text{MgO}$. However, at the maximum $\text{B}_2\text{O}_3$ content of 25 mol\%, the CR remained high (approximately 30 N) and was virtually unaffected by the $\text{MgO}$ for $\text{CaO}$ substitution. This insensitivity was mirrored when $\text{B}_2\text{O}_3$ content was varied in the presence of 15 mol\% $\text{MgO}$. Collectively, these results strongly suggest that the strengthening mechanisms induced by the addition of boron and the addition of magnesium represent two distinct and independent effects.

The CR is plotted as a function of RID in Fig.~\ref{fig:CR}c. As expected from the weak variation of RID and the large variation of CR across the composition range, we observe only a very weak dependence of CR on RID, in contrast to previous reports~\cite{Sellappan13, Rouxel_2021}. Interestingly, the slope of this dependence is negative, which is unexpected in the light of presently accepted trends. For completeness, data for the same commercial glass compositions (LS, SLS, and ABS) investigated by~\cite{Kato10} have been added to the plot (black markers). For these three glasses, a clear positive correlation between CR and RID is observed, conforming to the accepted ideas. These opposing trends are summarized in Fig.~\ref{fig:CR}d, where arrows indicate the direction of change in crack resistance. These observations reinforce the view that the correlation between CR and RID postulated eg in Ref.~\cite{Rouxel15} is only partial, as previously noted~\cite{Januchta19}.

\subsection{Indentation cross-sections and shear bands} %
Plastic flow under indentation was investigated as a function of glass composition using the bonded interface technique~\cite{Hagan80, Gross18}: indentation over a pre-existing crack near the crack tip, followed by full crack opening, provided cross-sections of 1 kgf indents for a wide range of compositions.

\begin{figure}[] %
\centering
\includegraphics[scale=0.64]{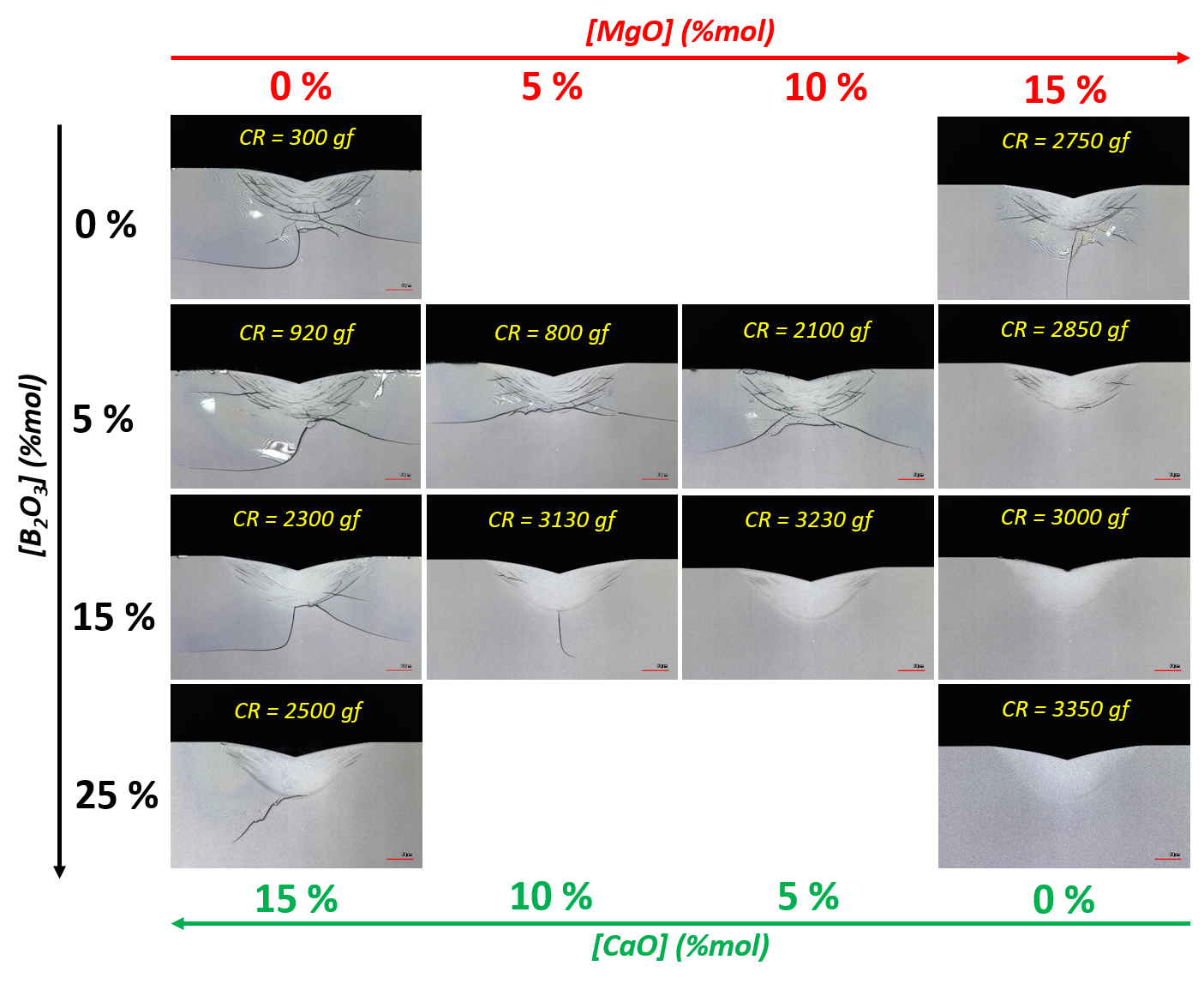}
\caption{Sum up of all cross-sections through 1 kgf Vickers indents}
\label{fig:PlasticFlow4}
\end{figure}

Figure~\ref{fig:PlasticFlow4} presents the results for both series of compositions, revealing a gradual evolution in the cross-section morphologies. The CAS glass (boron- and magnesium-free) exhibits pronounced shear localization, manifested as a fairly regular network of darker lines within the plastic zone. As $\text{B}_2\text{O}_3$ gradually substitutes for $\text{SiO}_2$ (from top to bottom), these shear faults progressively disappear and are replaced by a homogeneous area with a lighter grey shade. This transformation first appears in the central part of the plastic zone and subsequently extends towards the edges. A similar transition is observed when $\text{Mg}^{2+}$ is substituted for $\text{Ca}^{2+}$ (from left to right). In the high boron, high magnesium composition, the plastic zone, which is easily discernible by its lighter gray shade, appears completely homogeneous. Remarkably, this evolution closely parallels the progressive increase of the crack resistance reported in the previous section, from approximately 300 gf for CAS to 3350 gf for the 25 mol\% $\text{B}_2\text{O}_3$ and 15 mol\% $\text{Mg}^{2+}$ composition.

\begin{figure}[] %
\centering
\includegraphics[scale=0.68]{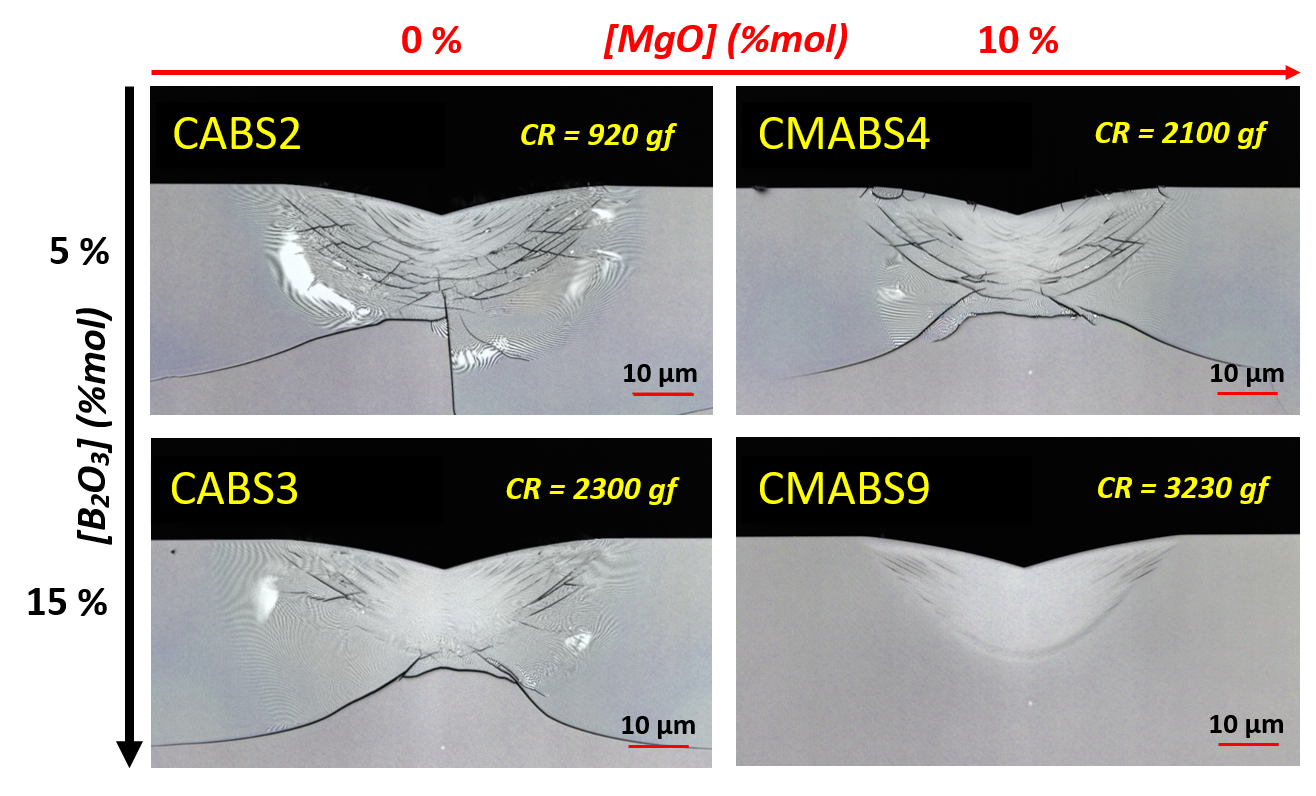}
\caption{Cross-sections through 1 kgf Vickers indents in (a) CABS2, (b) CMABS4, (c) CABS3 and (d) CMABS9.}
\label{fig:PlasticFlow2}
\end{figure}

Figure~\ref{fig:PlasticFlow2} displays cross-sections at 1 kgf for four mixed network modifier compositions, providing a clearer visualization of the distribution of shear faulting within the plastic zone. In the magnesium-free composition with 5 mol\% $\text{B}_2\text{O}_3$ (CABS2), shear faults are observed across the entire plastic zone, exhibiting the concentric distribution of shear bands under indentation and the symmetric sets of secondary 'spiral' bands characteristic of normal silicate glasses~\cite{Hagan80} or BMGs~\cite{Chen10}. This composition exhibits a low CR of approximately 900 gf, and both median and lateral cracks are visible. Upon the addition of 10 mol\% $\text{MgO}$ (CMABS4), the CR doubles, and the central part of the deformation zone contains fewer visible shear faults, while the edges show similar shear patterns to the CABS2 composition. A similar trend is observed for the CABS glass with 15 mol\% $\text{B}_2\text{O}_3$ (CABS3), where the central part of the indent is shear-fault free, yet the edges of the deformation zone still display a high density of shear bands/faults. Finally, with the addition of both boron and magnesium (CMABS9), the central part of the plastic zone becomes smoother, minimal shear localization is discernible on the edges and the lateral crack has disappeared. In this glass, the CR exceeds 3000 gf.

Finally, the commercial compositions studied in~\cite{Kato10} have also been investigated, and their cross-sectional views for 1~kgf indents are provided in the Supplemental Material (Fig.~S3) along with their CR taken from~\cite{Kato10}, confirming that the correlation between CR and shear banding also holds for these compositions.

These observations extend the work of Gross \emph{et al.}~\cite{Gross18} by demonstrating that CR systematically increases as shear faults become less pronounced, a behavior observed across all compositions studied.
\subsection{Roughness measurements of the plastically deformed region} \label{roughness} %
\begin{figure}[H] %
\centering
\includegraphics[width=0.8\textwidth]{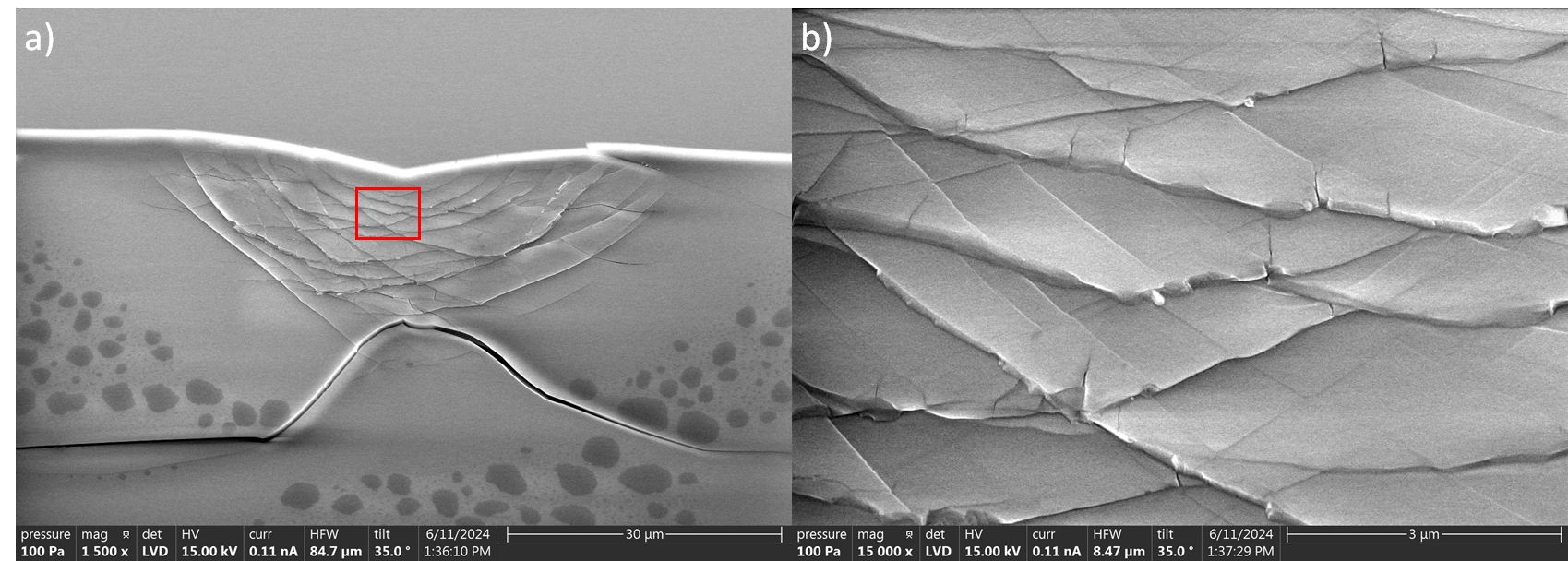}
c)\includegraphics[scale=0.35]{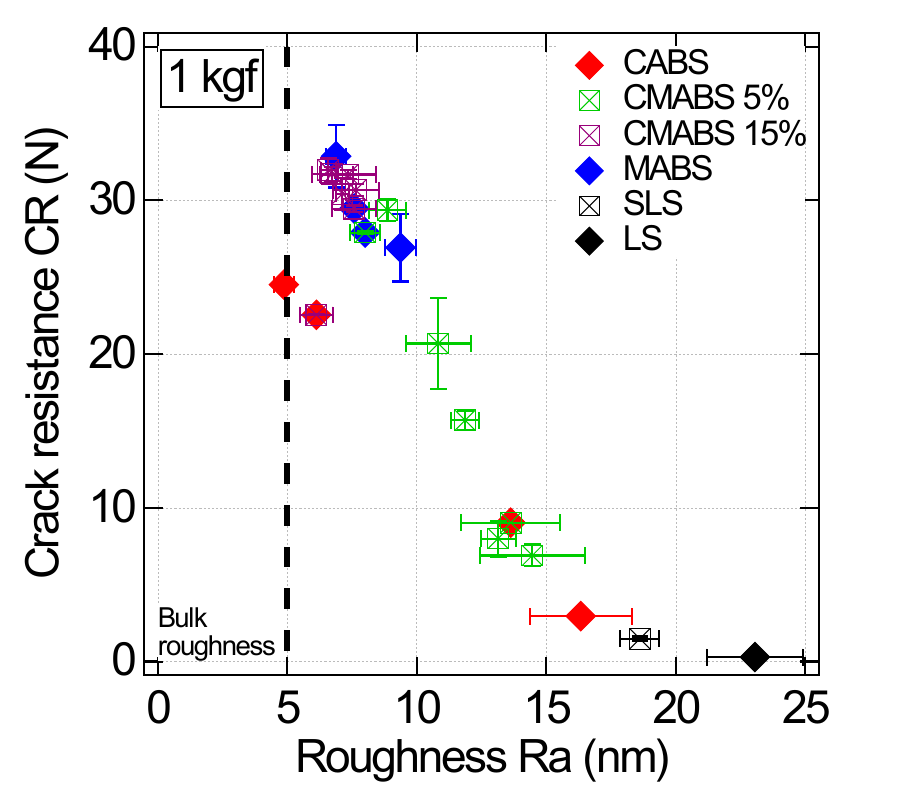}
d)\includegraphics[scale=0.35]{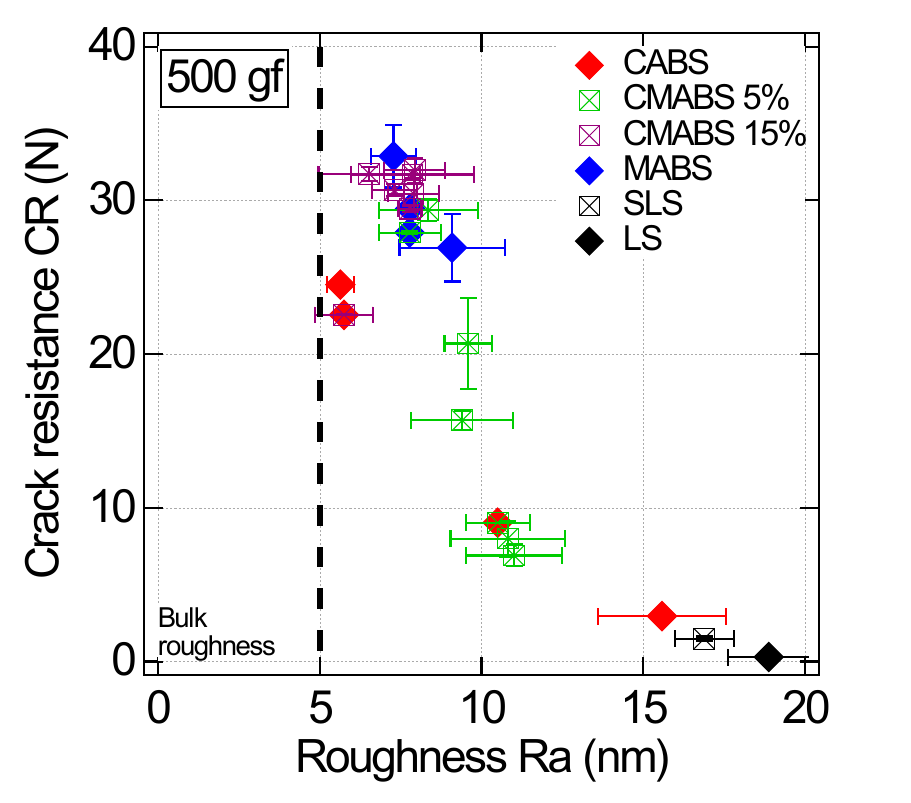}
e)\includegraphics[width=0.8\textwidth]{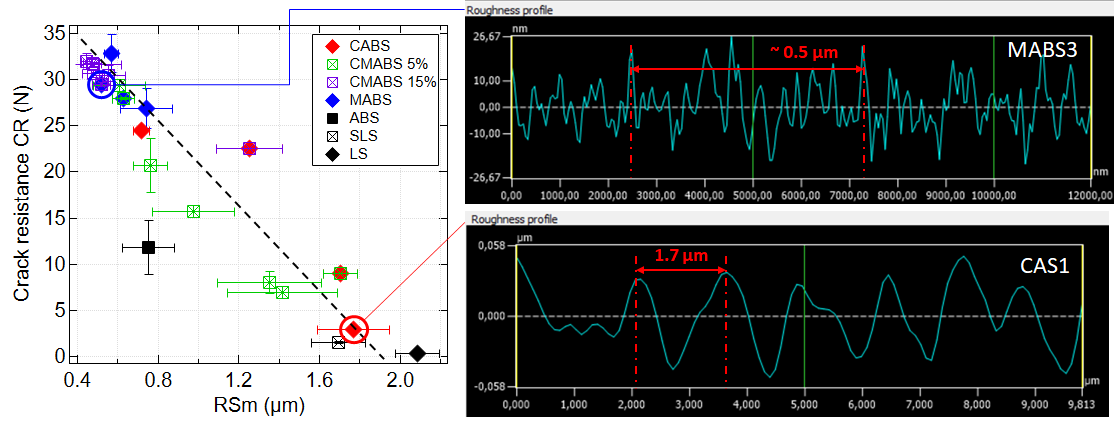}
\caption{(a) Scanning Electron image of 1kgf Vickers indentation cross-section in soda-lime silicate glass and (b) magnified image on shear bands from the cross-section. The sample has been tilted to 35° to observe shear bands from the bottom. (c-d) Relationship between crack resistance ($CR$) and roughness ($R_a$) for Vickers indentation cross-sections at 1 kgf (c) and 500 gf (d) including some commercial compositions (black markers) from Kato et al.~\cite{Kato10}. Vertical dashed line is determined as the bulk roughness. (e) Crack resistance as a function of the average spacing RS$_m$ between shear bands from Vickers indentation cross-sections at 1kgf for all glass compositions.}
\label{fig:Roughness}
\end{figure}

In the bonded interface method, shear bands intersect the original crack faces as slip steps~\cite{Schuh07} (Fig.~\ref{fig:Roughness} a and (zoom) b). As a result, in our experiments, the characteristics of the shear bands affect the corrugation of the cross-section. To quantify shear band activity under loading, we measured the surface roughness of the cross-section using standard topography parameters: roughness amplitude $R_a$ and characteristic lateral size RS$_m$. These measurements were performed on the central region beneath the indenter using laser scanning microscopy. Multi-lines used to measure roughness were drawn vertically onto the center of the deformation zone where shear bands intersect each other horizontally.

Figure~\ref{fig:Roughness}c illustrates the relationship between CR and the mean roughness $R_a$, measured from cross-sections of 1~kgf Vickers indents. This correlation indicates that more crack-resistant glasses, such as the high-boron compositions, exhibit lower roughness over the center of the plastic zone, suggesting less pronounced shear localization. Conversely, compositions with low crack resistance display higher roughness, i.e., more significant shear bands, as exemplified by the lead silicate glass.

We also investigated the effect of indentation load. Since shear bands initiate on the edge of the plastic zone, by geometrical invariance, a shear band that initiates under higher load develops a larger slip~\cite{Chen10}. Consequently, we find a larger surface roughness at the bottom of the plastic zone for a 1~kgf indent than for 500~gf (Figure~\ref{fig:Roughness}d). The ratio of approximately 1.2 in $R_a$ between the two loads is consistent with the $\sqrt{2}$ scaling expected from the self-similarity of the indentation process.

Finally, we have also calculated the characteristic correlation distance RS$_m$ in the roughness profiles, which is expected to reflect the spacing between shear bands. Stronger shear bands associate larger slip and increased spacing resulting in a decrease of crack resistance with RS$_m$ (Figure~\ref{fig:Roughness}e).

\subsection{Structural rearrangements and shear deformation mechanisms} %

To corroborate these macroscopic observations and relate them to the intrinsic structural features of the glass, we employed molecular dynamics (MD) -- or more precisely molecular statics -- simulations. MD is a well-established tool for probing the mechanical response of model amorphous materials.

For generic amorphous systems, MD studies have shown that shear localization initiates through the percolation of shear transformation zones (STZs) along planes of maximum shear stress~\cite{Shi05}, often via a vortex-like mechanism~\cite{Sopu17}. This process leads to structural rejuvenation within the shear band, which subsequently stabilizes the flow~\cite{Barbot20}. Moreover, MD investigations have demonstrated that the resulting failure mode—brittle localization versus ductile flow—is fundamentally governed by the thermodynamic stability of the glass~\cite{Richard20}. In highly stable glasses, yielding occurs as a sharp phase transition with a characteristic stress overshoot, whereas in poorly annealed glasses, it manifests as a smooth crossover to plastic flow~\cite{Ozawa18}.

More specifically, for silicate glasses, other MD studies have investigated their mechanical properties with particular focus on the relationship between network topology, plastic flow~\cite{Mantisi16, Molnar16} and rupture behavior~\cite{Yuan14, Molnar17}. More recently, significant attention has been devoted to borosilicate glasses~\cite{Deng18, Liu20} and their network rearrangement mechanisms. Under tensile loading, it was found that plastic flow promotes the concentration of non-bridging oxygens (NBOs) through bond-switching, with fracture initiating preferentially within these NBO-rich, mechanically weakest regions~\cite{Lee20, Lee21}.

\begin{figure}[H] %
\centering
\includegraphics[width=.75\textwidth]{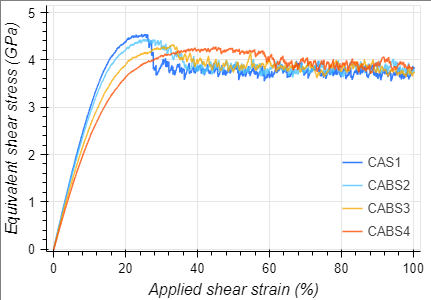}
\caption{Stress-strain curves in shear for Ca alumino silicate glasses with increasing boron content.}
\label{fig:stress_strain_MD}
\end{figure}

Here we have employed large scale molecular dynamics simulations using the SHIK potential~\cite{shih_new_2021} to investigate the mechanical behavior of generic borosilicate glasses. More detail on the simulation technique can be found in the Supplemental Material, section S~II. The resulting stress–strain curves under pure shear were calculated as in our previous work~\cite{Molnar17}. The results are shown in Figure~\ref{fig:stress_strain_MD}. The addition of boron leaves the flow stress largely unchanged but progressively lowers and broadens the yield-stress peak, thereby reducing the amplitude of strain softening. These combined effects are expected to decrease the susceptibility of the glass to shear banding. Indeed, the localization of mechanical deformation requires a reduction in either strength or stiffness. Compositions exhibiting weaker softening are therefore less prone to shear-band formation, as their shear resistance is only weakly affected by deformation-induced structural transformations. As a result, the shear strength is only marginally influenced, preventing any significant reduction. Shear localization is thus inhibited, leading to more homogeneous deformation, in agreement with the experimental observations.

\section{Discussion} %

There is a longstanding paradox in the fracture behavior of silicate glasses. Classical models of indentation cracking~\cite{Lawn77, Lawn80, Chiang82, Yoffe1982, Feng07} describe fracture in terms of three intrinsic material parameters: elasticity (Young’s modulus $E$), plastic yielding (hardness $H$), and fracture resistance (toughness $K_{IC}$). Yet glasses with very similar values of $E$, $H$, and $K_{IC}$ can display crack resistances that differ by more than an order of magnitude~\cite{Kato10, Sellappan13, Barthel20}. This discrepancy has increasingly directed attention toward a deformation mechanism specific to silicate glasses: irreversible volumetric strain, or densification~\cite{Ernsberger68}, which accompanies the more conventional volume-conserving shear plastic flow~\cite{Peter70, Arora79} during indentation.

Densification is thought to influence crack resistance by modifying the residual stress field generated upon unloading. In this context, the analytic indentation model proposed by Yoffe~\cite{Yoffe1982} was indeed able to account for a densification-induced reduction of tensile residual stresses, and thus for a reduction of crack initiation. Because direct measurements of densification remain experimentally challenging~\cite{RosalesSosa25}, it has often been inferred from more accessible proxies, most commonly the recovery of indentation depth at $0.9,T_g$ (RID)~\cite{Yoshida05}. Several studies have indeed reported strong correlations between crack resistance and RID~\cite{Kato10, Sellappan13, Rouxel15, Rouxel_2021}, reinforcing the view that densification plays a central role in glass cracking.

However, this interpretation has recently been questioned. The emphasis placed on densification and its proxies such as RID or Poisson’s ratio has been challenged~\cite{Januchta19,Barthel20}, and Yoffe’s model itself has been shown to provide only a crude description of the true indentation stress field~\cite{Davis20, RosalesSosa25}. These developments suggest that while densification contributes to crack resistance, its role should be more nuanced than previously assumed: additional mechanisms or stress-field complexities must be considered to fully resolve the paradox of glass fracture.

With the present series of glasses, we find no significant correlation between crack resistance and either Poisson’s ratio or RID, confirming that neither the composition dependence of elastic properties nor the permanent volumetric deformation modes can be regarded as the primary factors governing crack initiation under indentation. In contrast, by extending earlier approaches~\cite{Hagan80, Gross18}, we observe a broad and robust relationship between crack resistance and the morphology of shear bands in the region where median/radial cracks initiate: glasses exhibiting high crack resistance display homogeneous permanent deformation within the plastic zone.

Note that although plastic flow in silicate glasses is often associated with shear banding, homogeneous irreversible shear deformation has clearly been demonstrated in pillar compression experiments on pure amorphous silica~\cite{Kermouche16}. As composition varies, we observe that shear bands become increasingly pronounced as crack resistance decreases. More precisely, as already noted by Gross et al.~\cite{Gross18}, glasses with greater crack resistance are found to accommodate deformation through a denser network of fine shear bands, implying smaller slip amplitude per band and effectively delaying or preventing the evolution of these bands into critical shear faults. 

Because in the present experiments,  shear band formation emerges as a key factor affecting crack resistance in silicate glasses, we have leveraged the surface steps left by shear bands at intersecting planes~\cite{Schuh07} to quantify shear band activity. We have found that indeed the roughness of indent cross-sections in crack initiation regions correlates strongly with crack resistance. The robustness of these observations is underscored by their consistency across two glass series, whether strengthened through Mg-for-Ca or B-for-Si substitution, despite the distinct underlying microscopic mechanisms involved.

This picture is consistent with a broad framework for glassy materials. In many amorphous solids deformed below their glass transition temperature, the dominant mode of plastic deformation is indeed shear bands, which concentrate plastic flow within narrow zones of intense shear strain whether in glassy polymers ~\cite{Wu76, Bucknall77, Argon13} or in BMG~\cite{Argon79, Schuh07}. The transition from a stable shear band that accommodates plastic strain to a rapidly propagating crack is triggered by intense localized flow, which may drive severe structural evolution within the band~\cite{Li2002Nanometre, Schuh07, Nicolas18}.

Consequently, to reduce shear localization and fracture in bulk metallic glasses, a variety of strategies have been developed. Reducing sample size can induce a brittle to ductile transition by limiting the extent of shear band development~\cite{Toennies14}. For macroscopic samples, most approaches fall under the general concept of diffusing localization, including the introduction of structural heterogeneity~\cite{Das05}, which promotes the nucleation, branching, and intersection of multiple shear bands, the incorporation of nanocrystalline or dendritic inclusions~\cite{Hofmann08}, which can nucleate and arrest shear bands but may reduce strength, or even the design of the material for shear induced crystallization in the bands, which adds a strain hardening mechanism~\cite{Chen06} thereby mitigating catastrophic localization and enhancing toughness.

The picture is more or less the same for glassy polymers. The morphology of shear bands governs the resulting macroscopic fracture behavior~\cite{Wu76}. At high strain rates or low temperatures, deformation typically leads to the formation of individual, coarse 'slip' bands~\cite{Bucknall77, Wu76}. Once these coarse bands extend across the specimen, they serve as brittle failure initiators~\cite{Friedrich83, Argon13}. In contrast, at low strain rates or elevated temperatures, deformation can be accommodated by fine bands organized within broader, diffuse shear zones~\cite{Wu76, Bowden70, Bucknall77}. This diffuse mode enables the material to sustain large plastic strains, ultimately producing ductile fracture. As in BMG, a common strengthening mechanism resorts to inclusions with contrasted mechanical properties: soft nanoparticles, for example, can trigger and arrest dilation bands, a specific form of shear bands, with strong enhancement of the fracture toughness~\cite{Lazzeri95}.

The strong anticorrelation between crack resistance and shear band formation in silicate glasses thus reinforces the broader framework linking plastic flow localization to fracture behavior in silicate glasses. A fully consistent analysis of crack initiation begins at the continuum scale with indentation mechanics. For silicate glasses, this framework may be provided, for example, by Yoffe’s model~\cite{Yoffe1982}, by the more comprehensive formulation of Chiang, Marshall, and Evans (CME)~\cite{Chiang1982}, or by more recent finite-element approaches based on constitutive relations~\cite{Davis20,Barthel20}. From the resulting stress and strain fields, crack-initiation criteria can be postulated and their predictions compared with experimental observations. As an illustration, CME typically invoked a statistical distribution of pre-existing flaws to describe crack initiation: with this approach, a cracking threshold depending solely on the elastic modulus $E$, the plastic yielding or hardness $H$, and the fracture toughness $K_{Ic}$ emerges. However, CME already recognized that the ultimate origin of median and lateral cracks in glasses most likely lied in flaws associated with discrete shear faults, as documented in the contemporary literature~\cite{Hagan80,Lawn1980} and described extensively in the present work. In addition, these early studies emphasized that the major flaws preferentially developed at the intersections of shear faults or flow lines. Notably, analogous mechanisms were reported contemporaneously in other amorphous solids, including the nucleation of cracks at intersecting flow lines in glassy polymers such as polystyrene~\cite{Wu76} and poly(methyl methacrylate)~\cite{Ritter88}. This broader perspective shifts the emphasis from the yield threshold alone to developed shear-flow behavior as a controlling factor in crack initiation, offering a promising route to resolve the glass fracture paradox.

Once posited that resistance to cracking is fundamentally governed by the propensity of the amorphous network to either localize or diffuse shear deformation, the composition dependance can be reconsidered. In “normal” glasses, such as soda-lime silicates, non-bridging oxygens provide pathways of weaker ionic bonding that readily facilitate unstable shear faulting, thereby acting as precursors to extensive crack systems. In contrast, increasing the boron (B$_2$O$_3$) content promotes the formation of trigonal boron (III) units and lead to a high density of finer, less damaging shear bands. A similar principle applies to the substitution of calcium by magnesium: owing to the high field strength of Mg, the network remains highly connected and largely free of non-bridging oxygens, effectively suppressing the weak links that promote the formation of large, deleterious shear faults~\cite{Gross18,Fry23}. The resistance to localization is tightly related to the plastic shear flow properties and the associated structural variations and provide a direct and general basis to the notions of 'self-adaptive'~\cite{Januchta17a, Osada20, Ke20} or 'flexible and compliant'~\cite{Lee20,Wondraczek22} networks.

Unfortunately, despite extensive investigations of metallic glasses and glassy polymers, the relationship between structure, dynamics, and shear-flow localization in amorphous materials remains elusive, limiting a comprehensive understanding of the brittle-to-ductile transition~\cite{Tanguy21}. Beyond the well-known (but as seen here only partial) correlation with Poisson’s ratio~\cite{Lewandowski05}, qualitatively different mechanisms have been suggested, notably through possible connections to the $\beta$ relaxation~\cite{Qiao14}. From a modeling standpoint, the transition from sparse plasticity to persistent shear banding, driven for example by the STZ–vortex mechanism~\cite{Sopu17}, offers a perspective which may help rationalize the findings of topological constraint approaches and the associated flexible-to-rigid transition~\cite{Bauchy16}. More recently, in generic amorphous systems, a random critical point has been identified that separates these regimes~\cite{Berthier24}, with, for instance, brittle failure governed by the state of the melt relative to the fragile-to-strong transition~\cite{Atila25}. However, as these calculations predominantly concern relatively fragile networks in the Angell sense, the present experimental perspective underscores the need for comparable theoretical and modeling efforts focused on more open, covalent networks \emph{i.e.} stronger glasses, before a truly universal theory of the brittle-to-ductile transition in amorphous material is reached.

\section{Conclusion} %
In silicate glasses, the experimentally observed variations in crack resistance are far larger than can be accounted for by the measured changes in elastic, plastic, or fracture-propagation (toughness) properties alone. Instead, we demonstrate a strong correlation between indentation-induced cracking and shear-band formation, providing compelling evidence that the degree of plastic shear-flow localization is a key controlling factor in crack initiation. This conclusion places silicate glasses in line with bulk metallic glasses and glassy polymers. The underlying mechanisms, and their dependence on glass composition and structure, warrant more intensive investigation to achieve a comprehensive understanding of fracture in amorphous materials. In practice, strategies aimed at mitigating brittleness in silicate glasses should primarily focus on controlling and diffusing shear localization.

\begin{acknowledgments} %
This work was funded by Agence Nationale de la Recherche, project GaLAaD (project n° ANR-20-CE08-0002). For the purpose of open access, the authors have applied a CC-BY public copyright licence to any Author Accepted Manuscript version arising from this submission. We thank Ludovic Olanier and Jean-Claude Mancer for help with the development of the experimental set-up.
\end{acknowledgments}

\bibliography{Structure_of_silicates,Mechanics,Micromechanics_amorphous,Simulation_refs}

\end{document}



\title{Supplemental material for : "Fracture initiation in silicate glasses via a universal shear localization mechanism"} %



\author{Matthieu Bourguignon}
\affiliation{Soft Matter Sciences and Engineering, ESPCI Paris, PSL University, CNRS, Sorbonne University, 75005 Paris, France.}

\author{Gustavo Alberto Rosales-Sosa}
\author{Yoshinari Kato}
\affiliation{Nippon Electric Glass, 7-1, Seiran 2-Chome, Otsu, 520-8639, Shiga, Japan.}

\author{Bruno Bresson}
\affiliation{Soft Matter Sciences and Engineering, ESPCI Paris, PSL University, CNRS, Sorbonne University, 75005 Paris, France.}

\author{Hikaru Ikeda}
\author{Shingo Nakane}
\affiliation{Nippon Electric Glass, 7-1, Seiran 2-Chome, Otsu, 520-8639, Shiga, Japan.}

\author{Gergely Molnár}
\affiliation{CNRS, INSA Lyon, LaMCoS, UMR5259, 69621 Villeurbanne, France}

\author{Hiroki Yamazaki}
\affiliation{Nippon Electric Glass, 7-1, Seiran 2-Chome, Otsu, 520-8639, Shiga, Japan.}

\author{Etienne Barthel}
\affiliation{Soft Matter Sciences and Engineering, ESPCI Paris, PSL University, CNRS, Sorbonne University, 75005 Paris, France.}

\date{\today}


\maketitle


\section{Experimental procedures}\label{sec:experimental_procedures} %

\subsection{Sample preparation}\label{sec:sample_preparation}
Batches were mixed to yield 500 g of glass using industrial grade raw materials: silicon oxide $\text{SiO}_2$ (99.9\%), boric acid $\text{H}_3\text{BO}_3$ (56.8 wt\%), aluminum oxide $\text{Al}_2\text{O}_3$ (99.9 wt\%), calcium carbonate $\text{CaCO}_3$ (55.5 wt\%) and magnesium oxide $\text{MgO}$ (99.3 wt\%). After batches were prepared, small amounts (0.1 wt\% of the total batch) of fining agent $\text{SnO}_2$ were added to allow the removal of gas bubbles from the melt. For each batch, weighted raw materials were first melted in a 300 cc Pt-Rh crucible for 15h at a temperature of 1500 °C in an electric furnace and stirred one first time using a platinum rod to improve the homogeneity of the melt. After letting melt overnight (around 15h), the batch materials were stirred periodically and melted again at T$_m$ + 30 °C for 3h. An approximation of their glass transition temperature T$_g$ was determined by using the Facstage\textsuperscript{\textregistered} and Interglad\textsuperscript{\textregistered} softwares, based on the viscosity calculation of each melt. Melted glasses were poured onto a carbon plate and placed in an electric furnace to cool slowly. They were annealed at T$_g$ + 30~°C for 1h and then cooled at 3 °C/min up to room temperature to reduce residual thermal stresses. All obtained glasses were transparent except the composition $15\text{MgO} \cdot 15\text{Al}_2\text{O}_3 \cdot 25\text{B}_2\text{O}_3 \cdot 45\text{SiO}_2$ which showed a "bluish" transparent color after annealing probably due to phase separation. Each annealed glass was then cut and processed into several different samples. The glass transition temperature was precisely determined by a dilatometer (see next section). Finally, samples of the glasses were heated up in an annealer at 5 °C/min up to T$_g$ + 30 °C, held for 30 min, cooled at 3 °C/min down to T$_g$ - 150 °C and then cooled at 5°C/min down to room temperature to reduce residual mechanical stresses from post-processing.

\begin{table}[]
\caption{Single alkaline earth aluminoborosilicate glass compositions}
\label{table:SMG}
\begin{ruledtabular}
\begin{tabular}{l c c c c c}
Glass & SiO$_2$ & B$_2$O$_3$ & Al$_2$O$_3$ & CaO & MgO\\
& (mol \%) & (mol \%) & (mol \%) & (mol \%) & (mol \%)\\
\hline
CAS1 & 70 & - & 15 & 15 & -\\
CABS2 & 65 & 5 & 15 & 15 & -\\
CABS3 & 55 & 15 & 15 & 15 & -\\
CABS4 & 45 & 25 & 15 & 15 & -\\
\\
MAS1 & 70 & - & 15 & - & 15\\
MABS2 & 65 & 5 & 15 & - & 15\\
MABS3 & 55 & 15 & 15 & - & 15\\
MABS4 & 45 & 25 & 15 & - & 15\\
\end{tabular}
\end{ruledtabular}
\end{table}

\begin{table}[H]
\caption{Dual alkaline earth aluminoborosilicate glass compositions.}
\label{table:DMG}
\begin{ruledtabular}
\begin{tabular}{l c c c c c}
Glass & SiO$_2$ & B$_2$O$_3$ & Al$_2$O$_3$ & CaO & MgO\\
& (mol \%) & (mol \%) & (mol \%) & (mol \%) & (mol \%)\\
\hline
CMABS1 & 65 & 5 & 15 & 12.5 & 2.5\\
CMABS2 & 65 & 5 & 15 & 10 & 5\\
CMABS3 & 65 & 5 & 15 & 7.5 & 7.5\\
CMABS4 & 65 & 5 & 15 & 5 & 10\\
CMABS5 & 65 & 5 & 15 & 2.5 & 12.5\\
\\
CMABS6 & 55 & 15 & 15 & 12.5 & 2.5\\
CMABS7 & 55 & 15 & 15 & 10 & 5\\
CMABS8 & 55 & 15 & 15 & 7.5 & 7.5\\
CMABS9 & 55 & 15 & 15 & 5 & 10\\
CMABS10 & 55 & 15 & 15 & 2.5 & 12.5\\
\end{tabular}
\end{ruledtabular}
\end{table}

\noindent {\large \textbf{Measurement of glass properties}}

\vspace{0.2cm}

The glass transition temperature $T_g$ and coefficient of thermal expansion $CTE$ were estimated with thermomechanical analysis (TMA) by using a Bruker AXS TD5000SA machine (differential dilatometer) under He gas flow. The maximum temperature that can be reached is around 900°C and a heating rate of 3°C/min was applied. Those measurements have been performed by a technician from NEG company.

Density was determined using Archimedes' method (water displacement) with an uncertainty of $\pm$ 0.0002 g/cm$^3$. Young's modulus ($E$) and shear modulus ($G$) were measured by resonance method using the ASTM standard issued under the fixed designation E1875-08, with 40 $\times$ 20 $\times$ 2 mm$^3$ samples. A Japan Techno-Plus JE-RT3 with Young's modulus (JR-RT) and Shear Modulus (JG) modules were used to estimate these two elastic moduli. Each sample was first gold coated for 2 min. The frequency at which the measured intensity is highest is defined as the resonance frequency $f$. From this value, we can calculate Young's modulus $E$ and shear modulus $G$ from the sample dimensions with the following equations:

\begin{equation}
\centering
E=0.9465\cdot\frac{Mf^2}{w}\cdot\left(\frac{L}{t}\right)^3\cdot\left(1+6.59\left(\frac{t}{L}\right)^2\right)
\label{eqn:Young_modulus}
\end{equation}

\begin{equation}
\centering
G=3.933\cdot\frac{MLf^2}{wt}\cdot\left(\frac{s+1/s}{4s-2.52s^2+0.21s^6}\right)
\label{eqn:Shear_modulus}
\end{equation}

where $L$ is the length of the sample, $t$ is the thickness, $w$ is the width, $M$ is the weight, and $s=t/w$. The bulk modulus ($K$) and the Poisson ratio ($\nu$) were then calculated using the standard relations:

\begin{equation}
\centering
K=\frac{EG}{(9G-3E)}
\label{eqn:Bulk_modulus}
\end{equation}

\begin{equation}
\centering
\nu=\frac{E}{2G}-1
\label{eqn:Poisson_ratio}
\end{equation}

To determine Vickers hardness, 30 indents were made at 100 gf and a dwelling time of 15~s with a microhardness tester (MXT50, Matsuzawa Seiki Corp., Japan). Diagonals for each indent were measured and the Vickers hardness ($H_V$) values were then calculated using the following equation:

\begin{equation}
\centering
H_V=\frac{2\sin{(\frac{\theta}{2})}\cdot F}{D^2} =\frac{1.8544\cdot F}{D^2}
\label{eqn:Hardness}
\end{equation}

where F is the applied load, $\theta$ is the 136° angle between opposite faces of the Vickers tip, and D is the diagonal length of the indent. The hardness unit is HV (or kgf/mm). To convert HV to GPa, we have to multiply by 0.009807.\\

\noindent {\large \textbf{Plastic flow characterization}}

\vspace{0.2cm}

Cross-sections of 500 gf and 1 kgf Vickers indents were prepared by using Peter's method \cite{Peter70} which has been used later by Hagan \cite{Hagan80} and more recently by Gross \cite{Gross18}. This technique involves indenting across a pre-existing crack introduced into a 25 $\times$ 5 $\times$ 1 mm$^3$ glass specimen by performing a line of dozen indentations at 1 kgf. The crack is slowly propagated with a 3-point bending setup until it reaches 1 mm in length from each side of the indents line. When the load is released, a part of the crack becomes invisible. Indentations at several loads were then made along the crack at different positions. The distance between two indents must be at least 100 µm to prevent the stress field of the previous indent from affecting the indentation pattern of the next one. The two faces of the glass specimen are finally separated by fully propagating the pre-existing crack with the same 3-point bending setup.\\

\noindent {\large \textbf{Laser Scanning Microscope (LSM)}}

\vspace{0.2cm}

The indentation cross-sections from both sides have been observed at the Nippon Electric Glass Co.,Ltd. with a VK-X250K/X260K Laser Scanning Microscope from KEYENCE company. The observation was performed with the highest magnification (X28k) and a single scan mode was used. Samples were fixed vertically on the holder with double-face tape to observe their edges where the plastically deformed region of the indentation cross-section is located.\\

\noindent {\large \textbf{Roughness and spacing measurements}}

\vspace{0.2cm}

The roughness of the plastically deformed region under indents was determined by using the multi-line roughness measurement from the MultiFileAnalyzer software (VK-H1XME, KEYENCE). This function sets multiple measurement lines parallel to an initial line given by the user and calculates the roughness $Ra$ of each line with

\begin{equation}
Ra=\frac{1}{l_r}\int_{0}^{l_r} \lvert Z(x)-\overline{Z} \rvert \,dx
\label{eqn:Ra}
\end{equation}

Similarly, the characteristic length between roughness features was measured from

\begin{equation}
Rsm=\frac{1}{m}\sum_{i=1}^{m} Xsi
\label{eqn:Rsm}
\end{equation}

Concerning the $Rsm$ value, Xsi is the length of a single profile element. The peaks (valley) will be treated as noise and considered a part of the preceding valley (peak) if the height (depth) is less than 10\% of the maximum height or the length is less than 1\% of the segment length.\\

\noindent {\large \textbf{Scanning Electron Microscopy (SEM)}}

\vspace{0.2cm}

The indentation cross-sections on soda-lime silicate glass were observed in ESPCI with a Quattro Environmental Scanning Electron Microscope (ESEM - ThermoFisher\textsuperscript{\textregistered}). Observations were conducted at room temperature and in a low vacuum mode under a 100 Pa atmosphere. This mode allows the observation of the sample without gold coating.\\

\subsection{Mechanical and thermal characterization}

\begin{figure}[]
\centering
a)\includegraphics[width=0.465\textwidth]{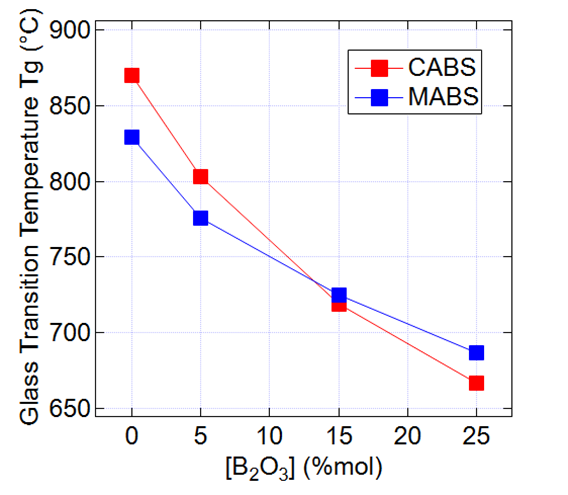}
b)\includegraphics[width=0.47\textwidth]{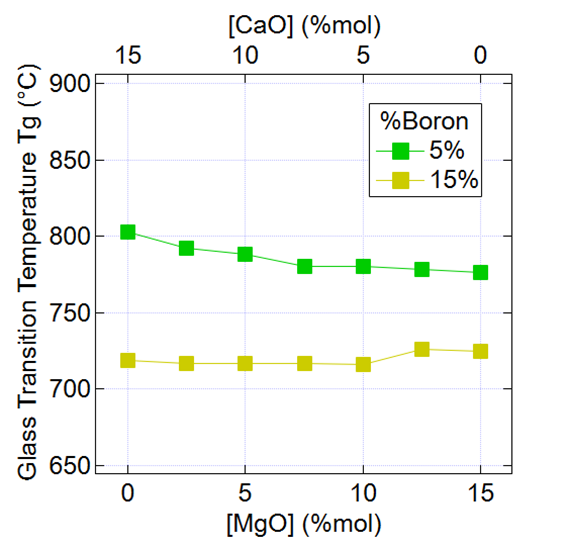}
\caption{Glass transition temperature $T_g$ of (a) single alkaline earth ABS glasses as a function of B$_2$O$_3$ content and (b) mixed alkaline earth ABS glasses as a function of MgO/CaO content. Error bars are smaller than the markers' size.}
\label{fig:Tg_CABS_CMABS_MABS}
\end{figure}

\begin{figure}[]
\centering
a)\includegraphics[width=0.48\textwidth]{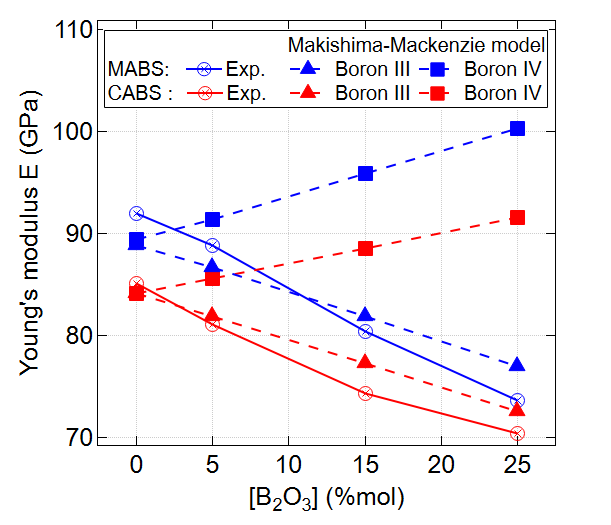}
b)\includegraphics[width=0.47\textwidth]{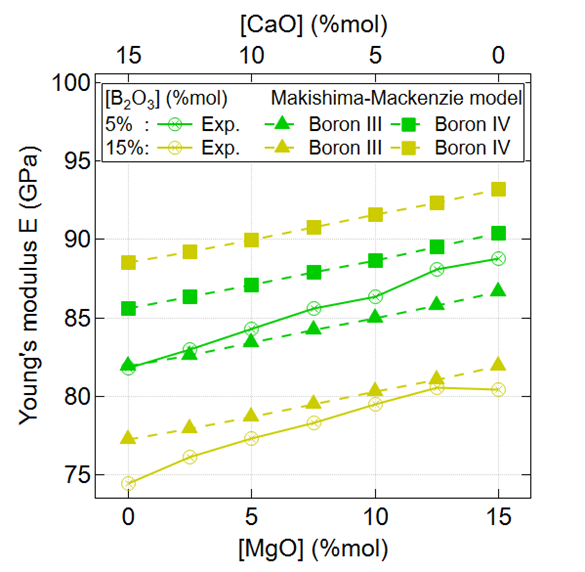}
\caption{Comparison of Young's modulus obtained experimentally with Makishima-Mackenzie model~\cite{Makishima73} depending on the boron coordination for (a) single alkaline earth ABS glasses and (b) mixed alkaline earth ABS glasses.}
\label{fig:Comparision_E_MM_model}
\end{figure}

Results of $\rho$, T$_g$, $E$, $G$, $K$, $\nu$, H$_v$ and $\sigma_y$ are shown in \textbf{Table \ref{tab:DataSMG}} and \textbf{\ref{tab:DataDMG}}. For all glasses (single or mixed alkaline earth network modifiers compositions), density decreases as B$_2$O$_3$ or MgO increases. Concerning the thermal characterization, $T_g$ decreases with the addition of B$_2$O$_3$ (Fig.~\ref{fig:Tg_CABS_CMABS_MABS}). Indeed, substituting a four-coordinated silica with a three-coordinated boron tends to gradually depolymerize the network and decrease the glass transition temperature. For a composition with 5\% B$_2$O$_3$, $T_g$ slightly decreases as MgO increases but it remains unchanged with 15\% B$_2$O$_3$. The values of $E$, $G$, $K$, and $H_v$ decrease with increasing B$_2$O$_3$ but increase (slightly concerning elastic moduli) with increasing MgO. This evolution of $E$ is consistent with the predictions of the Makishima-MacKenzie model~\cite{Makishima73} assuming a 100\% B(III) speciation, as expected for these compositions (Fig.~\ref{fig:Comparision_E_MM_model}).

On the other hand, the Poisson's ratio experiences minimal variation with the addition of boron, and remains generally unaffected by the substitution of calcium for magnesium.

\begin{table}[]
\caption{Density ($\rho$), Glass transition temperature (T$_g$), Young's modulus ($E$), Shear modulus ($G$), Bulk modulus ($K$), Poisson's ratio ($\nu$), Vickers hardness ($H_{\mathrm{V}}$), Yield stress ($\sigma_{\mathrm{y}}$), Crack resistance ($CR$) and Recovery indentation depth ($RID$) of single alkaline earth network modifier ABS glasses.}
\label{tab:DataSMG}
\begin{ruledtabular}
\begin{tabular}{l c c c c c c c c c c}
Glass & $\rho$ & $T_{\mathrm{g}}$ & $E$ & $G$ & $K$ & $\nu$ & $H_{\mathrm{V}}$ & $\sigma_{\mathrm{y}}$ & $CR$ & $RID$\\
& (g/cm$^3$) & (°C) & (GPa) & (GPa) & (GPa) & & (GPa) & (GPa) & (N) & (\%)\\
\hline
CAS1 & 2.52 & 870 & 85.1 & 35.1 & 49.6 & 0.21 & 7.1 & 3.5 & 3 & 30\\
CABS2 & 2.49 & 803 & 81.0 & 33.3 & 47.7 & 0.22 & 6.6 & 3.4 & 9 & 33\\
CABS3 & 2.45 & 719 & 74.4 & 30.1 & 46.9 & 0.24 & 6.1 & 3.1 & 23 & 29\\
CABS4 & 2.42 & 667 & 70.4 & 28.1 & 47.0 & 0.25 & 5.8 & 2.7 & 25 & 26\\
\\
MAS1 & 2.48 & 829 & 92.0 & 37.8 & 54.2 & 0.22 & 7.8 & 3.8 & 27 & 30\\
MABS2 & 2.46 & 776 & 88.8 & 36.4 & 52.8 & 0.22 & 7.7 & 3.7 & 28 & 26\\
MABS3 & 2.41 & 725 & 80.4 & 32.7 & 49.5 & 0.23 & 7.0 & 3.3 & 29 & 25\\
MABS4 & 2.37 & 687 & 73.6 & 29.8 & 46.3 & 0.24 & 6.5 & 3.1 & 33 & 27\\
\end{tabular}
\end{ruledtabular}
\begin{flushleft}
Experimental uncertainties are as follows: $\rho$: $\pm$0.01 g/cm$^3$; T$_g$: $\pm$2 °C; $E$, $K$ and $G$: $\pm$0.1 GPa; $\nu$: $\pm$0.01; H$_v$: $\pm$0.1 GPa; $\sigma_y$: $\pm$0.1 GPa; $CR$: $\pm$1 N; $RID$: $\pm$1 $\%$
\end{flushleft}
\end{table}

\begin{table}[]
\caption{Density ($\rho$), Glass transition temperature (T$_g$), Young's modulus ($E$), Shear modulus ($G$), Bulk modulus ($K$), Poisson's ratio ($\nu$), Vickers hardness ($H_{\mathrm{V}}$), Yield stress ($\sigma_{\mathrm{y}}$), Crack resistance ($CR$) and Recovery indentation depth ($RID$) of mixed alkaline earth network modifiers ABS glasses.}
\label{tab:DataDMG}
\begin{ruledtabular}
\begin{tabular}{l c c c c c c c c c c}
Glass & $\rho$ & $T_{\mathrm{g}}$ & $E$ & $G$ & $K$ & $\nu$ & $H_{\mathrm{V}}$ & $\sigma_{\mathrm{y}}$ & $CR$ & $RID$\\
& (g/cm$^3$) & (°C) & (GPa) & (GPa) & (GPa) & & (GPa) & (GPa) & (N) & (\%)\\
\hline
CMABS1 & 2.49 & 792 & 83.0 & 34.0 & 49.4 & 0.22 & 6.8 & 3.5 & 7 & 31\\
CMABS2 & 2.48 & 788 & 84.3 & 34.5 & 50.1 & 0.22 & 6.9 & 3.6 & 8 & 29\\
CMABS3 & 2.47 & 780 & 85.6 & 35.1 & 50.7 & 0.22 & 7.2 & 3.6 & 16 & 30\\
CMABS4 & 2.47 & 780 & 86.3 & 35.4 & 51.2 & 0.22 & 7.3 & 3.7 & 21 & 28\\
CMABS5 & 2.46 & 778 & 88.1 & 36.1 & 52.4 & 0.22 & 7.6 & 3.7 & 29 & 28\\
\\
CMABS6 & 2.44 & 717 & 76.1 & 30.9 & 47.6 & 0.23 & 6.2 & 3.2 & 30 & 29\\
CMABS7 & 2.44 & 717 & 77.3 & 31.4 & 47.8 & 0.23 & 6.3 & 3.2 & 31 & 29\\
CMABS8 & 2.43 & 717 & 78.3 & 31.8 & 48.5 & 0.23 & 6.4 & 3.2 & 32 & 26\\
CMABS9 & 2.42 & 716 & 79.5 & 32.3 & 49.3 & 0.23 & 6.7 & 3.3 & 32 & 27\\
CMABS10 & 2.42 & 716 & 80.5 & 32.7 & 49.9 & 0.23 & 6.9 & 3.3 & 32 & 27\\
\end{tabular}
\end{ruledtabular}
\begin{flushleft}
Experimental uncertainties are as follows: $\rho$: $\pm$0.01 g/cm$^3$; T$_g$: $\pm$2 °C; $E$, $K$ and $G$: $\pm$0.1 GPa; $\nu$: $\pm$0.01; H$_v$: $\pm$0.1 GPa; $\sigma_y$: $\pm$0.1 GPa; $CR$: $\pm$1 N; $RID$: $\pm$1 $\%$
\end{flushleft}
\end{table}

\begin{figure}[H]
\centering
\includegraphics[scale=0.55]{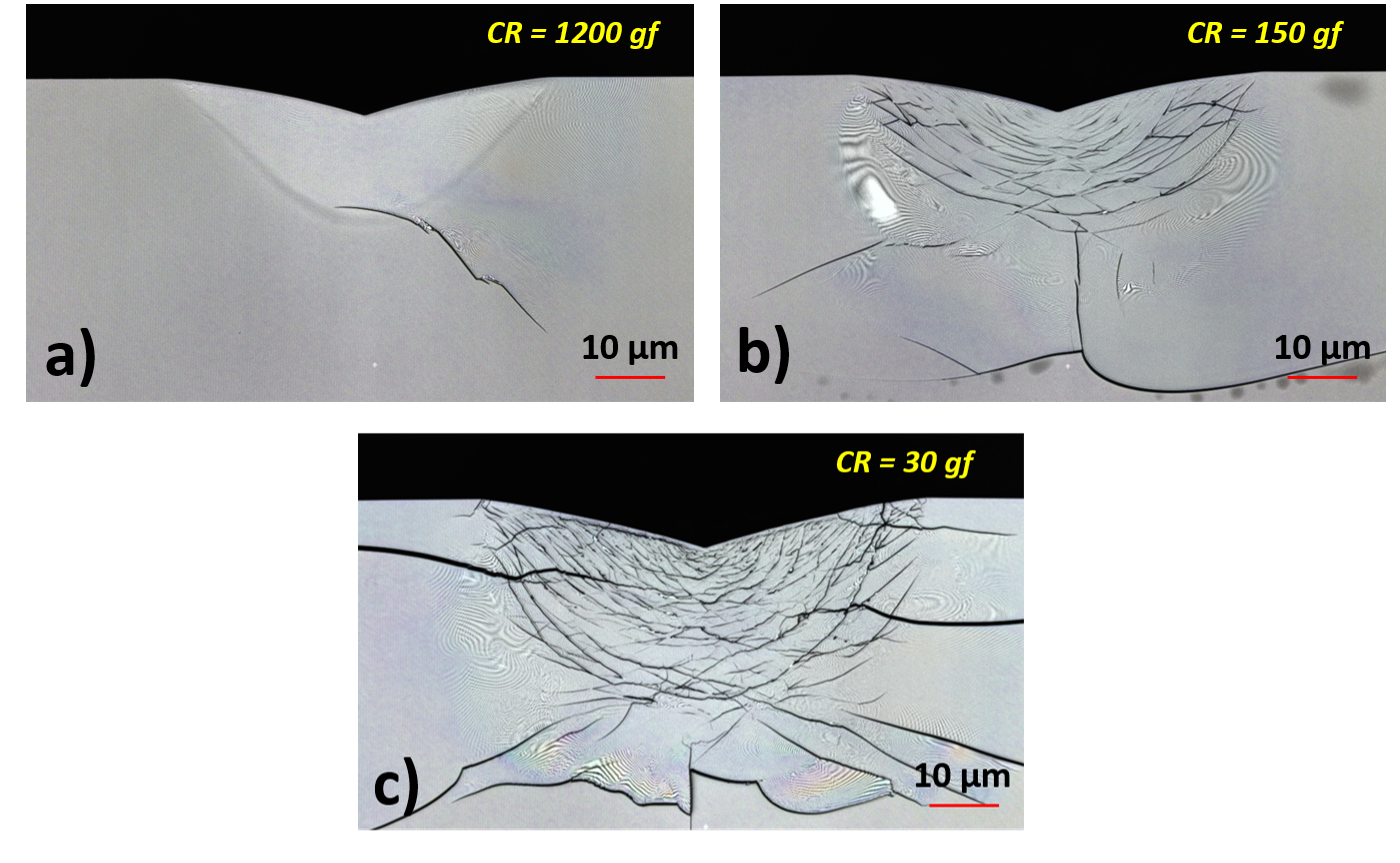}
\caption{Cross-sections through 1 kgf Vickers indents in (a) aluminoborosilicate, (b) soda-lime silicate and (c) lead silicate.}
\label{fig:PlasticFlow3}
\end{figure}

Their crack resistance values for the three commercial glasses LS, SLS and ABS studied in~\cite{Kato10} are taken from the same paper, but they have not been measured again for this work. The ABS glass (\textbf{Fig. \ref{fig:PlasticFlow3} (a)}) doesn't show any shear localization within the deformation zone. However, this glass is more sensitive to cracking than a 25\% B$_2$O$_3$ glass composition (CABS4 for instance) whose deformation zone doesn't show any shear pattern as well. On the other hand, soda-lime silicate glass (\textbf{Fig. \ref{fig:PlasticFlow3} (b)}) and lead silicate glass(\textbf{Fig. \ref{fig:PlasticFlow3} (c)}) exhibit significant shear faulting over all their deformation zones and also have a very high cracking resistance.

\section{Molecular Dynamics (Statics) Simulation procedure}\label{sec:simulation_details}
\setcounter{table}{0}
\renewcommand{\thetable}{B\arabic{table}}

\subsection{Potential function}
We adopted the SHIK potential \cite{sundararaman_new_2018} \cite{shih_new_2021} developed by Kob and co-workers as the force field model for molecular dynamics simulations. In this potential, the long-range Coulombic interactions are evaluated by the Wolf summation method \cite{wolf_exact_1999}, while the short-range interactions employ the Buckingham functional form \cite{buckingham_classical_1938}.
\begin{equation}
U_{ij}(r_{ij})=A_{ij}\exp(-B_{ij}r_{ij})-\frac{C_{ij}}{r_{ij}^6}+\frac{D_{ij}}{r_{ij}^{24}}+U_{ij}^{W}(r_{ij})
\label{eqn:line_1}
\end{equation}

Where,

\begin{equation}
U_{ij}^{W}(r_{ij})=z_{i}z_{j}\bigg[\frac{1}{r_{ij}}-\frac{1}{r_{cut}^{W}}+\frac{(r_{ij}-r_{cut}^{W})}{(r_{cut}^{W})^2}\bigg]
\end{equation}

The short-range interactions were truncated at 8 \AA, and the long-range Coulombic interactions were truncated at $r_{cut}^{W}$ = 10 \AA. To maintain continuity of both the potential and forces, the following smoothing function $G_{ij}(r_{ij})$ was multiplied:

\begin{equation}
G_{ij}(r_{ij})=\exp\bigg(-\frac{\gamma^2}{(r_{ij}-r_{cut})^2}\bigg)
\end{equation}

where the width of the smoothing function $\gamma$ is 0.3 \AA, and $r_{cut}$ is the short-range cutoff distance. The charges $z$ of Si, Al, and Ca are fixed, while the charge of O adopts a value that varies depending on the glass composition. In the reference \cite{sundararaman_new_2018}, $\gamma$ = 0.2 \AA\ was adopted for smoothing the potential function, but in this study, we used $\gamma$ = 0.3 \AA. This is to suppress numerical oscillations in the stress-strain curve during unloading in deformation simulations and obtain a smoother response. We verified the effect of changing $\gamma$ on Young's modulus and the main features of the stress-strain curve and confirmed that no significant difference occurred in the physical property values. Therefore, we judged that the adoption of $\gamma$ = 0.3 \AA\ contributes to improving simulation stability without compromising physical validity.
Table~\ref{tab:MD_parameters} shows a list of the interaction parameters we employed.

\begin{table}[]
\centering
\caption{Short-range interaction parameters.}\label{tab:MD_parameters}
\vspace{1em} 
\begin{tabular}{c|c c c c}
\hline
i-j & A$_{ij}$ (eV) & B$_{ij}$ (\AA$^{-1}$) & C$_{ij}$ (eV$\cdot$\AA$^6$) & D$_{ij}$ (eV$\cdot$\AA$^{24}$)\\
\hline
Si-Si & 2798.0 & 4.4073 & 0.0 & 3423204 \\
Si-Ca & 77366.0 & 5.0770 & 0.0 & 16800 \\
Si-O & 23108.0 & 5.0979 & 139.70 & 66 \\
Al-Al & 1799.1 & 3.6778 & 100.0 & 16800 \\
Al-O & 21740.0 & 5.3054 & 65.815 & 66 \\
Ca-Ca & 21633.0 & 3.2562 & 0.0 & 16800 \\
Ca-O & 146905.0 & 5.6094 & 45.073 & 16800 \\
O-O & 1120.5 & 2.8927 & 26.132 & 16800 \\
\hline
\end{tabular}
\end{table}

\subsection{Glass model generation}
We prepared a rectangular model of 100×600×100 \AA~containing approximately 480,000 atoms with periodic boundary conditions to evaluate the shear bands generated during shear deformation of glass. An initial model with randomly placed atoms was generated, and structural equilibration was performed at 3000 K for 100 ps. Subsequently, the system was quenched to $10^{-5}$ K at a cooling rate of 10 K/ps.These processes were controlled by the NPT ensemble. After that, structural relaxation was performed in the NVT ensemble for 100 ps, followed by relaxation in the NVE ensemble for 100 ps. Through these processes, a quenched glass model was obtained \cite{molnar_sodium_2016}.

\subsection{Glass model deformation simulation}
We conducted simple shear deformation simulations of the glass under conditions without applying hydrostatic pressure to the model. The shear strain was applied by tilting the simulation box size in the xy plane. During this process, the dimensions of the simulation box were controlled at constant displacement steps to ensure that no hydrostatic pressure was applied to the model, i.e., the pressure in each axis satisfied $\sigma_{xx}=\sigma_{yy}=\sigma_{zz}=0$. After model size control, atomic positions were relaxed by minimizing the total potential energy of the system. This deformation process was repeated until the accumulated strain reached a specified value. The unloading process followed the same procedure.

\bibliography{Structure_of_silicates,Mechanics,Micromechanics_amorphous,Simulation_refs}